\documentclass[aps,showpacs]{revtex4}
\usepackage{amsbsy}
\usepackage{bbold}
\usepackage{graphicx,color}
\usepackage{amsfonts}
\usepackage{amsmath}
\usepackage{amssymb}
\usepackage{amsthm,bm}
\usepackage{dsfont}

\newcommand{\nn}{\nonumber}

\begin{document}
\title{Exciton ground-state energy with full hole warping structure}
\author{Roland Combescot$^{a}$ and Shiue-Yuan Shiau$^{b}$
}
\affiliation{$^{a}$
Laboratoire de Physique, Ecole Normale Sup\'erieure, PSL Universit\'e, 
Sorbonne Universit\'e, Paris Diderot Universit\'e, CNRS, 24 rue Lhomond, F-75005 Paris,
France.
\\$^{b}$Physics Division, National Center for Theoretical  Sciences, 10617,Taipei, Taiwan.
}
\date{Received \today}

\begin{abstract}

Most semiconductors, in particular III-V compounds, have a complex valence band structure near the band edge, 
due to degeneracy at the zone center. One peculiar feature is the warping of the electronic dispersion relations, which are not isotropic
even in the vicinity of the band edge. When the exciton, all important for the semiconductor optical
properties, is considered, this problem is usually handled by using some kind of angular averaging procedure, 
that would restore the isotropy of the hole effective dispersion relations. In the present paper, we consider the problem of the
exciton ground-state energy for semiconductors with zinc-blende crystal structure, and we solve it exactly
by a numerical treatment, taking fully into account the warping of the valence band. In the resulting four-dimensional 
problem, we first show exactly that the exciton ground state is fourfold degenerate. We then explore the ground-state
energy across the full range of allowed Luttinger parameters. We find that the correction due to warping may in principle
be quite large. However, for the semiconductors with available data for the band structure we have considered, the
correction turns out to be in the $10\% - 15\%$ range.

\end{abstract}
\maketitle

\section{INTRODUCTION}

Optical properties of semiconductors are of utmost interest both for fundamental and applied purposes \cite{cardona,vurga}.
With respect to these properties, excitonic excitations play a prominent role, especially in the visible light spectrum. 
Hence, knowing the exciton
ground-state properties in these compounds is of fundamental importance. However, this is in general not a simple matter theoretically,
since generically the valence band of the most relevant semiconductors is degenerate at the band edge.
In the standard case of semiconductors with cubic zinc-blende crystal structure, the states at the valence band edge
have a p-wave character which leads to a sixfold degeneracy, when spin is taken into account. When spin-orbit coupling is strong
enough to make a twofold degenerate split-off subband far enough in energy from the valence band edge, so that it becomes
irrelevant to the excitonic structure, one is left with a fourfold degeneracy. This is the situation we will consider.

Unfortunately, it has been long recognized that, even in the vicinity of the band edge, 
the resulting valence band dispersion relations are not isotropic, while
respecting the cubic symmetry of the crystal. As a result, the equal energy surfaces are warped, and this whole matter
makes the hole properties difficult to handle. This complexity is usually disposed of by taking an appropriate
angular average to go back to an isotropic situation. In the present paper, we will not make use of such a simplification, 
but rather explore the exciton ground-state energy in these zinc-blende semiconductors in full generality.

In the next section, we recall the hole effective kinetic energy Hamiltonian, standard for these compounds.
In the following section III, we write the Schr\"odinger equation for the exciton under the form of an integral equation
appropriate for our study. We show in section IV in full generality that, due to the cubic symmetry, the exciton ground
state has a fourfold degeneracy (omitting the twofold additional degeneracy coming from the conduction electronic spin). 
Finally, in section V we solve numerically our integral equation. We show that
in principle the warping leads to an increase of the exciton binding energy, which may be quite important. However,
in practice, for the tabulated semiconductors we have found, this increase is limited to a $10\% - 15 \%$ range.
Nevertheless, markedly higher values are possible, considering the imprecision of the known valence band parameters data,
or other semiconductors yet to be investigated.

\section{Formalism}\label{form}

In the semiconductor compounds we are considering, the bottom of the conduction band located at the zone center 
is non-degenerate (except for the spin degree of freedom), but the top of the valence band located at the zone center is, 
in the absence of spin-orbit coupling, threefold degenerate (forgetting the hole spin), with a p-wave-like
character for the corresponding wave functions. As the exciton wave function is predominantly made of electron and hole 
states in the vicinity of the zone center,
this degeneracy has to be taken into account in the exciton wave function. Taking spin into account, this threefold degeneracy becomes a
sixfold degeneracy. When spin-orbit coupling is taken into account, this degeneracy is lifted into a fourfold degeneracy, 
plus a twofold one. In standard compounds, these last two degenerate electronic states are shifted to energy low enough, 
so that they are irrelevant to the building of the exciton. This is the situation we will restrict ourselves to in the following. 
The four-dimensional hole subspace we now focus on has the
symmetry character of a $J=3/2$ angular momentum \cite{lutt}, resulting from the composition of the p-wave-like wave functions and of the $1/2$ spin. For the present purpose, it is more convenient to use the notations 
$|1\rangle \equiv | 3/2\rangle$, $|2\rangle \equiv | 1/2\rangle$, $|3\rangle \equiv | -1/2\rangle$
and $|4\rangle \equiv | -3/2\rangle$ to denote the standard fourfold hole basis. 

Following the approach of Luttinger and Kohn \cite{lutkoh,lutt} (with slightly different notations \cite{note1})
in the absence of magnetic field,
 the hole kinetic energy part $H_h({\bf k})$ of the exciton Hamiltonian is given by
\begin{eqnarray}\label{eqHtr}
H_h({\bf k}) = \frac{\hbar^2}{2m_0} \left( \begin{array}{cccc} G+\Delta & D & F & 0 
\\ D^{*} & G-\Delta & 0 & F \\ F^{*} & 0 & G-\Delta & -D \\ 0 & F^{*} & -D^{*} & G+\Delta \end{array} \right)
\equiv \frac{\hbar^2}{2m_0} (G\, \mathbb{1}+h_h)
\end{eqnarray}
where $m_0$ is the vacuum electron mass (providing a mass scale), and $G,\Delta ,D$ and $F$ are defined from the hole
wavevector ${\bf k}$ by
\begin{eqnarray}\label{eqdefGD}
G&=&\gamma_1 k^2 \hspace{15mm} \Delta = \gamma_2 (k^2-3k_z^2) \\ \nonumber
D&=&-2 \sqrt{3} \gamma_3 (k_x-ik_y) k_z \\
F&=&-\sqrt{3} \gamma_2 (k_x^2-k_y^2)+2i\sqrt{3} \gamma_3 k_x k_y \nonumber
\end{eqnarray}
where $x,y$ and $z$ are the cubic axes, and $\gamma_1$, $\gamma_2$ and $\gamma_3$ are the Luttinger \cite{lutt} constants
characterizing the semiconductor.

One may notice that, within a multiplicative factor, $h_h^2$ reduces to the unit matrix
\begin{eqnarray}\label{eqhh2}
\left( \frac{2m_0}{\hbar^2} H_h({\bf k})-G\, \mathbb{1} \right)^2=h_h^2= \lambda^2 \, \mathbb{1}
\end{eqnarray}
with
\begin{eqnarray}\label{eqD2F2}
\lambda^2 \equiv \Delta^2 + |D|^2 + |F|^2=4 \gamma_2^2 k^4 +12 (\gamma_3^2-\gamma_2^2)\left[k_x^2k_y^2+k_y^2k_z^2+k_z^2k_x^2\right]
\end{eqnarray}
so that the doubly degenerate eigenvalues $E_{\pm}({\bf k})$ of $H_h({\bf k})$ are given by the well-known result \cite{dkk}
\begin{eqnarray}\label{eqEwarp}
\frac{2m_0}{\hbar^2} E_{\pm}({\bf k}) = \gamma_1 k^2 \pm 2 \Big[ \gamma_2^2 k^4 +3 (\gamma_3^2-\gamma_2^2)\left[k_x^2k_y^2+k_y^2k_z^2+k_z^2k_x^2\right]\Big]^{1/2}
\end{eqnarray}
The square root on the right-hand side gives rise to the well-known warping, implying in general a departure from a dispersion
relation with simple spherical symmetry. This symmetry is recovered only in the particular case where $\gamma_2=\pm\gamma_3$.
In this specific case, one finds the standard heavy and light holes, with mass $m_H$ and $m_L$, related to our coefficients by
(assuming $\gamma_2>0$)
\begin{eqnarray}\label{eqmLmH}
\frac{m_0}{m_H}=\gamma_1-2\gamma_2\hspace{20mm} \frac{m_0}{m_L}=\gamma_1+2\gamma_2
\end{eqnarray}

In order to make sense as a hole dispersion relation, $E_{\pm}({\bf k})$ which gives the hole kinetic energy
should naturally be positive for any ${\bf k}$. This implies some limitations on the Luttinger coefficients, from the explicit expression
given by Eq.(\ref{eqEwarp}). Indeed, considering the negative branch $E_{-}({\bf k})$, this requires the square root on the right-hand side to be less than $\gamma_1 k^2$. If we consider first the case where $\gamma_3 < \gamma_2$, the square root is maximal when
the bracket multiplying $\gamma_3^2-\gamma_2^2$ is zero, which occurs when $k_x=k_y=0$, or $k_y=k_z=0$, or $k_z=k_x=0$,
that is, when ${\bf k}$ is on one of the cubic axes. In this case, the positivity condition is merely
\begin{eqnarray}\label{eqcondgam32}
\gamma_1 > 2 \gamma_2
\end{eqnarray}
In the opposite case where $\gamma_3 > \gamma_2$, in order to maximize the square root, we instead look for the
maximal value of the bracket multiplying $(\gamma_3^2-\gamma_2^2)$. For fixed $k$, it is found when $\pm k_x=\pm k_y=\pm k_z=k/\sqrt{3}$,
that is, when ${\bf k}$ is along one of the principal cubic diagonals. In this case, the square root is merely equal to $\gamma_3 k^2$,
and the positivity condition is (assuming $\gamma_3>0$)
\begin{eqnarray}\label{eqcondgam23}
\gamma_1 > 2 \gamma_3
\end{eqnarray}
In the particular case where $\gamma_3=\gamma_2$, these conditions merely reduce to the fact that $m_H$ should be positive,
which is physically obvious.

\section{Integral equation for excitonic eigenstates}\label{ieftee}

The Hamiltonian for the exciton is merely obtained by adding to the hole kinetic energy the electronic kinetic energy ${\bf p}_e^2/2m_e$,
where ${\bf p}_e$ is the conduction band electronic momentum, and $m_e$ the corresponding electronic band mass. We assume that
the electronic dispersion relation is isotropic \cite{ross}. 
The electronic spin does not bring any complication, except that naturally all the degeneracies are multiplied by two;
so, we do not indicate it explicitly. Finally, we have to include in the Hamiltonian the electron-hole
attractive Coulomb interaction $V({\bf r}_e-{\bf r}_h)$, where ${\bf r}_e$ and ${\bf r}_h$ are respectively the electron and hole position, with
$V({\bf r})=-e^2/(4\pi \epsilon_{sc} r)$ where $\epsilon_{sc}$ is the semiconductor dielectric constant. Since this interaction depends only
on the relative position of the hole and the electron, it is translationally invariant, so the excitonic momentum is conserved. In the present
paper, we restrict ourselves to the case where this momentum is zero, so that the electron and hole momentum are opposite 
${\bf p}_e=-{\bf p}_h=-\hbar {\bf k}$. Hence, the electron term merely adds a contribution $\hbar^2 k^2/2m_e$ to the hole kinetic energy
Eq.(\ref{eqHtr}), and we are back to a one-body problem. The potential term is the Coulomb interaction, and the kinetic energy is still given
by Eq.(\ref{eqHtr}), provided $\gamma_1$ is replaced by
\begin{eqnarray}\label{eqgam1bar}
{\bar \gamma}_1=\gamma_1+\frac{m_0}{m_e}
\end{eqnarray}

Considering the complexity of the hole kinetic energy, it is more convenient to write the excitonic Schr\"odinger equation in
momentum space, which leads to an integral equation. This is much easier to handle numerically than the
four-dimensional partial differential equation one would find if one worked in real space. We denote the ${\bf k}$ representation of the
four-components wave function as $a({\bf k}) \equiv \big( a_1({\bf k}),a_2({\bf k}),a_3({\bf k}),a_4({\bf k})\big)$.
This four-dimensional Schr\"odinger equation then reads
\begin{eqnarray}\label{eqexcSch}
{\bar H}_h({\bf k}) \, a({\bf k}) +  \int \frac{d{\bf k}'}{(2\pi)^3}\,V({\bf k}-{\bf k}')\, a({\bf k}') = E\, a({\bf k})
\end{eqnarray}
where $V({\bf q})=-e^2/(\epsilon_{sc} q^2)$ is the Fourier transform of the Coulomb interaction, and $E$
is the exciton energy to be solved for. ${\bar H}_h({\bf k})$ is just given by Eq.(\ref{eqHtr}), except that $\gamma_1$ has to
be replaced by ${\bar \gamma}_1$. Note that the Coulomb interaction is diagonal in our representation,
while ${\bar H}_h({\bf k})$ is not.

It is convenient to rewrite this equation so that only dimensionless quantities appear. The natural energy
unit is the Rydberg, so we write the exciton energy as $E=- {\bar E}\, m_0 (e^2/4\pi \epsilon_{sc})^2/(2 \hbar^2)$
where ${\bar E}$ is the dimensionless exciton binding energy. The corresponding natural length unit is the Bohr
radius, so we also go to a dimensionless wavevector ${\bar {\bf k}}$ defined by 
${\bf k}={\bar {\bf k}} \,m_0 (e^2/4\pi \epsilon_{sc})/\hbar^2$. Taking into account that ${\bar H}_h({\bf k})$
is a quadratic function of the wavevector, our Schr\"odinger equation becomes
\begin{eqnarray}\label{eqexcScha}
\big( {\bar H}_r({\bar {\bf k}})+{\bar E} \big) a({\bar {\bf k}}) 
= \frac{1}{\pi^2} \int d{\bar {\bf k}}' \frac{1}{({\bar {\bf k}}-{\bar {\bf k}}')^2}\, a({\bar {\bf k}}')
\end{eqnarray}
where we have also defined a reduced dimensionless hole kinetic energy ${\bar H}_r({\bar {\bf k}})=
(2m_0/\hbar^2){\bar H}_h({\bar {\bf k}})$. 

Furthermore, it is clear from Eq.(\ref{eqHtr}) and Eq.(\ref{eqgam1bar})
that the important parameters are the ratios $\gamma_2/{\bar \gamma}_1$ and $\gamma_3/{\bar \gamma}_1$.
Setting
\begin{eqnarray}\label{eqgambar}
{\bar \gamma_2} =\frac{\gamma_2}{{\bar \gamma}_1} \hspace{20mm} 
{\bar \gamma_3} =\frac{\gamma_3}{{\bar \gamma}_1}
\end{eqnarray}
it is convenient to have only these parameters appearing. This requires a further rescaling, obtained by setting
${\bar {\bf k}}=K/{{\bar \gamma}_1}$ and $ {\bar E}= {\bar \epsilon}/{{\bar \gamma}_1}$. This leads to
\begin{eqnarray}\label{eqexcSchb}
\big( (K^2+{\bar \epsilon})\mathbb{1}+{\bar h}_h({\bf K}) \big) a({\bf K}) 
= \frac{1}{\pi^2} \int d{\bf K}' \frac{1}{({\bf K}-{\bf K}')^2}\, a({\bf K}')
\end{eqnarray}
where ${\bar h}_h({\bf K})$ has exactly the same expression as $h_h({\bf k})$ in Eq.(\ref{eqHtr})
(with ${\bf k}$ replaced by ${\bf K}$) except that, in the explicit expressions for its elements given in
Eq.(\ref{eqdefGD}), one has to replace $\gamma_2$ and $\gamma_3$ by ${\bar \gamma_2}$ and
${\bar \gamma_3}$ respectively.

We note that, corresponding to our above request for the positivity of the hole kinetic energy, we have
to require that the kinetic energy for the exciton one-body problem is always positive.
Proceeding as in the above derivation of Eq.(\ref{eqcondgam32}) and Eq.(\ref{eqcondgam23}), we find
that our coefficients must satisfy
\begin{eqnarray}\label{eqcondgambar}
|{\bar \gamma_2}| < \frac{1}{2} \hspace{20mm} |{\bar \gamma_3}| < \frac{1}{2}
\end{eqnarray}

For numerical work, it is more convenient to apply to Eq.(\ref{eqexcSchb}) the inverse of the $4 \times 4$ matrix
$(K^2+{\bar \epsilon})\mathbb{1}+{\bar h}_h({\bf K})$. Since we have noted in Eq.(\ref{eqhh2}) 
and Eq.(\ref{eqD2F2}) that ${\bar h}_h^2=\lambda^2 \mathbb{1}$ is proportional to the unit matrix 
(with $\gamma_{2,3}$ and ${\bf k}$ replaced in $\lambda$ by ${\bar \gamma_{2,3}}$ and ${\bf K}$), this inverse is merely
$\big((K^2+{\bar \epsilon})\mathbb{1}-{\bar h}_h({\bf K})\big)/((K^2+{\bar \epsilon})^2-\lambda^2)$.
Hence Eq.(\ref{eqexcSchb}) becomes
\begin{eqnarray}\label{eqexcSchc}
 a({\bf K}) = \frac{(K^2+{\bar \epsilon})\mathbb{1}-{\bar h}_h({\bf K})}{(K^2+{\bar \epsilon})^2-\lambda^2}
\frac{1}{\pi^2} \int d{\bf K}' \frac{1}{({\bf K}-{\bf K}')^2}\, a({\bf K}')
\end{eqnarray}
Note that the conditions Eq.(\ref{eqcondgambar}) correspond to the requirement that the denominator
$(K^2+{\bar \epsilon})^2-\lambda^2$ is always positive, in particular for large $K$.

To conclude this section, it is useful to consider the particular case where our exciton problem reduces to
the case of the hydrogen atom. This corresponds merely to the case where ${\bar h}_h({\bf K})=0$ and 
the integral equation (\ref{eqexcSchc}) reduces to a one-dimensional equation 
\begin{eqnarray}\label{eqexcSchcH}
 a({\bf K}) = \frac{1}{\pi^2 (K^2+{\bar \epsilon})} \int d{\bf K}' \frac{1}{({\bf K}-{\bf K}')^2}\, a({\bf K}')
\end{eqnarray}
We readily know the solution which is the Fourier transform of the ground-state wave function $\exp(-r/a_0)$
of the hydrogen atom,
that is (omitting the unimportant constant prefactor) $a({\bf K})=1/(1+K^2)^2$. Taking into account that, with our
reduced units, the ground-state energy is merely ${\bar \epsilon}=1$, one can easily check analytically that
this expression of $a({\bf K})$ is indeed solution of the integral equation (\ref{eqexcSchcH}).

It is important to note that the explicit large $K$ behaviour $a({\bf K}) \sim 1/K^4$ of this solution, 
which insures the large $K$ convergence
of the integral on the right-hand side, is actually a generic feature of the general equation (\ref{eqexcSchc}).
Indeed, making use of this convergence consistently allows us to write in this large $K$ regime $1/({\bf K}-{\bf K}')^2
\simeq 1/K^2$ in the integral. Then, the resulting $1/K^2$ factor, together with the explicit prefactor of the integral,
indeed leads to $a({\bf K}) \sim 1/K^4$ for the large $K$ behaviour of $a({\bf K})$. This behaviour is naturally of 
particular interest for the numerical solution of this integral equation.

\section{Ground state degeneracy}\label{grstdeg}

One might think at first that Eq.(\ref{eqexcSchc}) gives for the exciton ground state a single non-degenerate solution, with a specific symmetry.
This is not what happens. Instead, we find that quite generally the exciton ground state, at zero total momentum, has an exact fourfold degeneracy (without taking into account the conduction electron spin degeneracy). This comes directly from the cubic symmetry of the crystal. This is fully
analogous to what would occur if we had a complete rotational invariance. In this last case, this would naturally
imply a full degeneracy, with the eigenstates having in this four-dimensional subspace an angular
dependence given by the spherical harmonics $Y_{\ell m}(\theta,\varphi)$ with $\ell=3/2$ and $m=(\pm 3/2,
\pm 1/2)$. In the present case, we have actually the lower cubic symmetry, but it is enough to similarly
insure the complete degeneracy of the ground state.

This result may at first look surprising if one naively considers the situation with spherical symmetry ${\bar \gamma_2}={\bar \gamma_3}$.
Since in this case one has heavy and light holes, one could think of two possible exciton states, made respectively with a heavy
or a light hole (each one being doubly degenerate). But this omits the fact that the Coulomb interaction is not diagonal in this heavy-light hole representation. Indeed, it has been shown \cite{monsean} that the proper treatment of this specific case leads to a fourfold
degeneracy. Here, we extend this result to the general case.

To see this in a more specific way, we have to consider the symmetry properties of the solutions. 
We rewrite the Schr\"odinger equation Eq.(\ref{eqexcSchb}) for the exciton, making
explicit the wave function and matrix components. It reads, with ($m,n)=1,2,3,4$

\begin{eqnarray}\label{eqexcSchm}
 \sum_{n=1}^{4} \big( (K^2+{\bar \epsilon}) \delta_{mn}+{\bar h}_{mn}({\bf K}) \big) a_n({\bf K}) 
= \frac{1}{\pi^2} \int d{\bf K}' \frac{1}{({\bf K}-{\bf K}')^2}\, a_m({\bf K}')
\end{eqnarray}
where ${\bar h}_{mn}({\bf K})$ are the matrix elements of ${\bar h}_h({\bf K})$
and $\delta_{mn}$ is the Kronecker symbol.
The fact that this equation satisfies time-reversal invariance implies that, if 
$a_{n}({\bf K})$ is solution of this equation, $(-1)^{n-1}a^*_{5-n}({\bf K})$ is also solution with the same energy.
This transformation can be understood from the spin-$3/2$ nature of our components. This property can be checked directly 
from the explicit expression of the hole kinetic energy terms ${\bar h}_{mn}({\bf K})$, given by Eq.(\ref{eqdefGD}).

Let us first consider the symmetry properties under a $\pi/2$ rotation $R_z$ around our quantization
axis $z$. Under this rotation, the component of the wave function $a_{n}({\bf K})$ becomes $a_{n}(R_z^{-1}({\bf K}))$, that is, 
$a_{n}(K_x,K_y,K_z)$ becomes $a_{n}(K_y,-K_x,K_z)$. For clarity, we denote as ${\bar a}_{n}$ the transform of $a_{n}$,
that is, ${\bar a}_{n}(K_x,K_y,K_z)=a_{n}(K_y,-K_x,K_z)$. Similarly, if we perform an additional $\pi/2$ rotation $R_z$
on the wave function, amounting to a total $\pi$ rotation, we end up with ${\bar{\bar a}}_{n}(K_x,K_y,K_z)=a_{n}(-K_x,-K_y,K_z)$.
Finally, an additional $\pi /2$ rotation leads to ${\bar{\bar{\bar a}}}_{n}(K_x,K_y,K_z)=a_{n}(-K_y,K_x,K_z)$.

However, if we want to rotate the state $ |a\rangle =  \sum_{n} a_n |n\rangle$ by a $\pi/2$ rotation $R_z$,
we have to take into account that our basis states $|3/2\rangle \equiv |1\rangle$,  $|1/2\rangle \equiv |2\rangle$,
$|-1/2\rangle \equiv |3\rangle$ and $|-3/2\rangle  \equiv |4\rangle$ change under 
rotation by acquiring a phase factor through the action of the rotation operator $e^{-i\pi J_z /2\hbar}$.
Disregarding the unimportant overall phase factor $e^{-3i\pi/4}$, this leads for the various components $a_n$ 
of our wave function to an additional phase factor $e^{i(n-1)\pi/2}=i^{n-1}$
coming from our basis states.

If we now make in Eq.(\ref{eqexcSchm}) the change of variables $K_x \to K_y$ and $K_y \to -K_x$
(and similarly for ${\bf K}'$ on the right-hand side), 
$a_{n}({\bf K})$ becomes ${\bar a}_{n}(\bf K)$ while we see from their explicit expressions Eq.(\ref{eqdefGD})
that $D({\bf K})$ becomes $i\,D({\bf K})$, $F({\bf K})$ becomes $-F({\bf K})$ and $\Delta ({\bf K})$
is unchanged. Comparing with the original equation (\ref{eqexcSchm}) before the change of variables,
one finds the following symmetry property: if $a_{n}$ is solution of Eq.(\ref{eqexcSchm}), 
then $i^{n-1}\, {\bar a}_{n}$ is also solution of this equation with the same energy. That is, if we have a solution,
the wave function obtained by a $\pi /2$ rotation is also a solution.
This physically corresponds to the fact that our Hamiltonian is invariant under a $\pi/2$ rotation $R_z$.
Naturally, repeating the $\pi/2$ rotation, we find that $(-1)^{n-1}\,{\bar{\bar a}}_{n}$ and $(-i)^{n-1}\,{\bar{\bar{\bar a}}}_{n}$
are also solutions.

This situation is analogous to the simple one-dimensional case where the Hamiltonian is invariant under the transformation $x \to -x$:
 if $\psi(x)$ is a solution of the Schr\"odinger equation, $\psi(-x)$ is also a solution with the same energy. 
So, we can build the two solutions $\psi_S(x)=\psi(x)+\psi(-x)$, and
$\psi_A(x)=\psi(x)-\psi(-x)$ which are respectively symmetric and antisymmetric under $x \to -x$.

In the same way, in our case, if $a_{n}({\bf K})$ is a solution of the Schr\"odinger equation (\ref{eqexcSchm}),
then by adding the solutions obtained by performing rotations around the $z$ axis by $\pi /2$, $\pi $ and $3\pi /2$ respectively,
we obtain a solution with the same energy which is invariant by $\pi /2$ rotation. This solution is explicitly
\begin{eqnarray}\label{eqasup1}
a^{(1)}_{n}=a_{n}+i^{n-1}\, {\bar a}_{n}+(-1)^{n-1}\,{\bar{\bar a}}_{n}+(-i)^{n-1}\,{\bar{\bar{\bar a}}}_{n}
\end{eqnarray} 
If we perform on this solution
the change of variables $K_x \to K_y$ and $K_y \to -K_x$, we find that $a^{(1)}_{n}$ is changed into $(-i)^{n-1}\,a^{(1)}_{n}$,
that is ${\bar a}^{(1)}_{n}=(-i)^{n-1}\,a^{(1)}_{n}$.

But, as possible solutions of Eq.(\ref{eqexcSchm}), we can more generally take solutions $a^{(p)}_{n}$, with $p=1,2,3,4$,
which transform as ${\bar a}^{(p)}_{n}=(-i)^{n-p}\,a^{(p)}_{n}$.
Generalizing the preceding case $p=1$, they are obtained explicitly from a solution $a_{n}({\bf K})$ by
\begin{eqnarray}\label{eqdefapn} 
a^{(p)}_{n}=a_{n}+i^{n-p}\, {\bar a}_{n}+(-1)^{n-p}\,{\bar{\bar a}}_{n}+(-i)^{n-p}\,{\bar{\bar{\bar a}}}_{n}
\end{eqnarray}
Performing the changes $K_x \to K_y$ and $K_y \to -K_x$, we check as for Eq.(\ref{eqasup1}) that they indeed
transform as we have indicated.
These are the appropriate generalizations to the present case of choosing solutions which are either even or
odd under $x \to -x$, when the Hamiltonian is invariant under this transformation $x \to -x$.
These four solutions are orthogonal since the scalar product of two different solutions is naturally unchanged when
one performs on ${\bf K}$ in the integral the change of variables $K_x \to K_y$ and $K_y \to -K_x$ 
corresponding to the $\pi /2$ rotation, while it is
changed due to the different way the two different solutions with $p \neq q$ transform under the rotation
\begin{eqnarray}\label{eqorth}
\sum_{n=1}^{4} \int d{\bf K} \,a^{(p)\,*}_{n}({\bf K})a^{(q)}_{n}({\bf K})=
\sum_{n=1}^{4} \int d{\bf K} \,{\bar a}^{(p)\,*}_{n}({\bf K}){\bar a}^{(q)}_{n}({\bf K})=
i^{q-p}\,\sum_{n=1}^{4} \int d{\bf K} \,a^{(p)\,*}_{n}({\bf K})a^{(q)}_{n}({\bf K})=0
\end{eqnarray}

In our one-dimensional case, if we start with an even solution $\psi(x)=\psi_S(x)$, the corresponding odd
solution $\psi_A(x)$ is found equal to zero. And conversely, if we start with $\psi_A(x)$, the corresponding
even solution is equal to zero. Similarly, in our case with $\pi /2$ rotation, one finds easily that if our starting
$a_{n}$ has a $a^{(p)}_{n}$ symmetry, the resulting solutions with different symmetry $a^{(q)}_{n}$, with
$q \neq p$, are identically equal to zero.

We note that the time-reversal counterpart of $a^{(1)}_{n}$, that is $(-1)^{n-1}a^{(1)*}_{5-n}$, transforms like $a^{(4)}_{n}$, and 
similarly $a^{(2)}_{n}$ and $a^{(3)}_{n}$ are linked in the same way. A solution and its time-reversed function can not be identical
because they have different symmetries. Hence, the time reversability automatically implies
that there is a twofold degeneracy for the exciton eigenstates (this is Kramer's degeneracy theorem with respect to hole states). 
In order to reach the conclusion that the
degeneracy is actually fourfold, we have to search for a further symmetry of the Hamiltonian.

For this purpose let us now consider the symmetry corresponding to a $\pi/2$ rotation $R_x$ around the $x$ axis.
With respect to the orbital variables, the situation is analogous to the one we had above with $R_z$:
$a_{n}({\bf K})$ becomes $a_{n}(R_x^{-1}({\bf K}))$, that is, 
$a_{n}(K_x,K_y,K_z)$ becomes ${\tilde a}_{n}(K_x,K_y,K_z)=a_{n}(K_x,K_z,-K_y)$.
However, in order to obtain the transform of our four-dimensional wave function, we again have naturally
to take into account the fact that our basis states are strongly modified under this rotation, that is, to calculate
$e^{-i\pi J_x /2\hbar}|n\rangle$. This is a standard calculation, that we will not detail here. 
It can be done for example by finding the evolution
of the matrix elements $f_n(\varphi)=\langle n| e^{-i \varphi J_x/\hbar }|m\rangle$ as a function of $\varphi$. For example
for $m=1$ one writes the first-order differential equations obtained by evaluating 
$\hbar \partial f_n(\varphi)/\partial \varphi=-i \langle n| J_x\,e^{-i \varphi J_x /\hbar}|1\rangle$ 
(for example $\partial f_1(\varphi)/\partial \varphi=(-i\sqrt{3}/2) f_2(\varphi)$ from the explicit expression of $J_x$) 
and one integrates these equations.

One obtains in this way the transform $|n\rangle _x \equiv e^{-i\pi J_x /2\hbar}|n\rangle$ of our basis states by the $R_x$ rotation as
\begin{eqnarray}\label{eqmatr}
|1\rangle _x &=& \frac{1}{2\sqrt{2}} \big [\,|1\rangle -i\,\sqrt{3}\,|2\rangle -\sqrt{3}\,|3\rangle +i\,|4\rangle \big ] \\ \nn
|2\rangle _x &=& \frac{1}{2\sqrt{2}} \big [-i\,\sqrt{3}\,|1\rangle -|2\rangle -i\,|3\rangle -\sqrt{3}\,|4\rangle \big ] \\ \nn
|3\rangle _x &=& \frac{1}{2\sqrt{2}} \big [-\sqrt{3}\,|1\rangle -i\,|2\rangle -|3\rangle -i\,\sqrt{3}\,|4\rangle \big ] \\ \nn
|4\rangle _x &=& \frac{1}{2\sqrt{2}} \big [i\,|1\rangle -\sqrt{3}\,|2\rangle -i\,\sqrt{3}\,|3\rangle +|4\rangle \big ] 
\end{eqnarray}
Hence, the solutions $ \sum_{n}a^{(p)}_{n}|n\rangle $ are transformed into 
$\sum_{m,n}{\tilde a}_{n}^{(p)}|m\rangle\langle m | n\rangle _x$, where the matrix elements $\langle m|n\rangle _x$
are given by Eq.(\ref{eqmatr}). These transformed solutions are naturally also solutions of the initial Schr\"odinger 
equation (\ref{eqexcSchm}). If the solutions $a^{(p)}_{n}$ are fourfold degenerate, this implies that the transformed solutions
are linear combinations of the $a^{(p)}_{n}$, that is, there exists a set of coefficients $\lambda^{(q)}_{p}$ such that
\begin{eqnarray}\label{eqident}
\sum_{n}{\tilde a}_{n}^{(q)}({\bf K}) \langle m | n\rangle _x=  \sum_{p} \lambda^{(q)}_{p} a^{(p)}_{m}({\bf K})
\end{eqnarray}
These simple linear relations between the solutions of Eq.(\ref{eqexcSchm}) are fairly remarkable, since the
solutions of Eq.(\ref{eqexcSchm}) are not expected to be simple. Since we have no analytical solutions of
Eq.(\ref{eqexcSchm}), we have no way to check them in the general case.

However, there is a limiting situation where we can solve analytically Eq.(\ref{eqexcSchm}) and check these
relations. This is the limit of large wavevectors ${\bf K}$, corresponding physically to short distances. In this case,
since on the right-hand side of Eq.(\ref{eqexcSchm}) the $a_{m}({\bf K'})$ have a limited range, we can write in the integral
$1/({\bf K-K'})^2 \simeq 1/{\bf K}^2$, which provides an explicit solution for $a_{n}({\bf K})$ in terms of 
the integrals $I_{n}= \int d{\bf K}\,a_{n}({\bf K})$, by merely inverting the matrix 
$K^2 \delta_{mn}+{\bar h}_{mn}({\bf K})$ as we have done in Eq.(\ref{eqexcSchc})
(${\bar \epsilon}$ is negligible in this large ${\bf K}$ limit). This gives explicitly
\begin{eqnarray}\label{eqexcSchcKgd}
 a_m({\bf K}) =  \sum_{n=1}^{4} \frac{K^2 \delta_{mn} -{\bar h}_{mn}({\bf K})}{K^4-\lambda^2}
\frac{I_{n}}{\pi^2 K^2}
\end{eqnarray}

If we take the particular case of the $ a^{(p)}_{n}$ solutions, we notice that, when we make 
in $I_{n}^{(p)}= \int d{\bf K}\,a_{n}^{(p)}({\bf K})$ the change of variables on ${\bf K}$
corresponding to the $R_z$ rotation, the integral is naturally unchanged. On the other hand, taking into account
that $a_{n}^{(p)}({\bf K})$ is transformed into ${\bar a}_{n}^{(p)}({\bf K})=(-i)^{n-p}\,a^{(p)}_{n}({\bf K})$ 
in this change of variables, we find that $I_{n}^{(p)}$ is multiplied by $(-i)^{n-p}$. Hence $I_{n}^{(p)}$
is zero unless $n=p$, so that $I_{n}^{(p)} \equiv I^{(p)}\,\delta_{n,p}$. Actually we can even choose to have
$I^{(p)}=1$ by taking an appropriate norm for the $ a^{(p)}_{n}$ solutions. Hence, in this case Eq.(\ref{eqexcSchcKgd})
simplifies into
\begin{eqnarray}\label{eqexcSchcKgda}
 a^{(p)}_m({\bf K}) = \frac{K^2 \delta_{mp} -{\bar h}_{mp}({\bf K})}{\pi^2 K^2(K^4-\lambda^2)}
\end{eqnarray}
In particular, one can check on these explicit expressions that the symmetry properties
${\bar a}^{(p)}_{n}=(-i)^{n-p}\,a^{(p)}_{n}$ are indeed satisfied.

Integrating Eq.(\ref{eqident}) over ${\bf K}$, and noting that $ \int d{\bf K}\,{\tilde a}_{n}^{(q)}({\bf K})=
\int d{\bf K}\,a_{n}^{(q)}({\bf K})=\delta_{n,q}$ by change of variables, we obtain the coefficients
$\lambda^{(q)}_{p}$ in Eq.(\ref{eqident}) as $\lambda^{(n)}_{m}=\langle m | n\rangle_x$, and in particular $\lambda^{(n)}_{n}=\langle n | n\rangle_x$. One can then substitute in Eq.(\ref{eqident}) the explicit solutions $a_{n}^{(q)}({\bf K})$ given by Eq.(\ref{eqexcSchcKgda}), and check after a simple but tedious calculation 
that it is satisfied. For example, taking the case $q=1$ and the component $m=1$, we have to check 
from Eq.(\ref{eqmatr}) that
\begin{eqnarray}\label{eqidenta}
{\tilde a}_{1}^{(1)}({\bf K})-i\sqrt{3}{\tilde a}_{2}^{(1)}({\bf K})-\sqrt{3}{\tilde a}_{3}^{(1)}({\bf K})
+i{\tilde a}_{4}^{(1)}({\bf K})=
a_{1}^{(1)}({\bf K})-i\sqrt{3}a_{1}^{(2)}({\bf K})-\sqrt{3}a_{1}^{(3)}({\bf K})+ia_{1}^{(4)}({\bf K})
\end{eqnarray}
where we have omitted the common denominator $2\sqrt{2}$ occuring in all the terms from Eq.(\ref{eqmatr}).
Using the definition ${\tilde a}_{n}(K_x,K_y,K_z)=a_{n}(K_x,K_z,-K_y)$, and noting that
${\tilde a}_{4}^{(1)}({\bf K})=a_{1}^{(4)}({\bf K})=0$, we are left with checking that
\begin{eqnarray}\label{eqcheck}
[K^2-{\bar \gamma}_2(K^2-3K_y^2)]+6i{\bar \gamma}_3 (K_x+iK_z)K_y-
[3{\bar \gamma}_2 (K_x^2-K_z^2) +6i{\bar \gamma}_3 K_xK_z]&=& \\ \nn
[K^2-{\bar \gamma}_2(K^2-3K_z^2)]-6i{\bar \gamma}_3 (K_x-iK_y)K_z&-&
[3{\bar \gamma}_2 (K_x^2-K_y^2) -6i{\bar \gamma}_3 K_xK_y]
\end{eqnarray}
where we have omitted the denominator $\pi^2 K^2(K^4-\lambda^2)$ common to all the terms.
We see that this equality Eq.(\ref{eqcheck}) is indeed satisfied, but in a nontrivial way.

It is now clear that the degeneracy of all the four solutions $a^{(p)}_{n}$ is necessary in order to satisfy such an
identity. Indeed, if on the contrary $a^{(1)}_{n}$ (together with its time-reversed conjugate
$a^{(4)}_{n}$) and $a^{(2)}_{n}$ (together with $a^{(3)}_{n}$) had different energies, the transformed
of $a^{(1)}_{n}$ by the rotation $R_x$ would have to be a linear combination of $a^{(1)}_{n}$ and $a^{(4)}_{n}$ only,
since the two other solutions had a different energy.
This would lead to relations analogous to Eq.(\ref{eqidenta}), but with only two solutions ($a^{(1)}$ and $a^{(4)}$)
on the right-hand side instead of four as in Eq.(\ref{eqidenta}). This is a much more stringent algebraic
requirement, that clearly can not be satisfied. Indeed, one can check in the large ${\bf K}$ limit
that the corresponding relation is algebraically incompatible with the explicit solution. This provides
an explicit proof that the four solutions $a^{(p)}_{n}$ are indeed degenerate \cite{remdegnum}.

\section{Numerical results}\label{numerics}

We numerically find the ground state energy (and the corresponding wave function)
by applying an iterative procedure, to this Schr\"odinger equation under the form written in Eq.(\ref{eqexcSchc}).
Let us call $A$ the linear operator that acts on the four-dimensional wave function $a({\bf K})$ 
on the right-hand side of Eq.(\ref{eqexcSchc}). 
Solving Eq.(\ref{eqexcSchc}) is equivalent to finding an eigenvector of $A$ with the eigenvalue $\Lambda=1$. 
In the case of a very large binding energy ${\bar \epsilon}$, the operator $A$ goes to zero and all its eigenvalues 
are quite small. Hence, none of them can be equal to $1$, and there is no state with very large binding energy, 
as expected. If we decrease ${\bar \epsilon}$, the largest positive eigenvalue $\Lambda_{\mathrm max}$ of $A$ 
will grow. When this eigenvalue of $A$ reaches $1$, we will have obtained the largest possible value for ${\bar \epsilon}$ corresponding to an eigenstate. In other words, we will have found the ground-state energy.

It is easy to obtain the largest eigenvalue of $A$ by applying iteratively $A$ to some convenient starting 
wave function $a({\bf K})$. A practical choice for the starting wave function is to take the first component
$a_1({\bf K})$ equal to the hydrogen atom ground-state wave function we have discussed above, and the other components
equal essentially to zero. Indeed, iterating $n$ times is equivalent to applying the operator
$A^n$ to $a({\bf K})$. But for large $n$ values, $A^n$ is dominated by its largest eigenvalue 
$(\Lambda_{\mathrm max})^n$ and is essentially equivalent to a projection on the corresponding eigenvector 
and multiplication by $(\Lambda_{\mathrm max})^n$. This allows us to identify conveniently $\Lambda_{\mathrm max}$
and the corresponding eigenvector. Actually this procedure works only if the spectrum of $A$ does not have nasty features, such as closely spaced largest and second largest eingenvalues, or large negative eigenvalues. 
Fortunately, in our case, this procedure happens to work quite nicely. We have found that, in practice,
typically 20 iterations, or less, gave already a satisfactory convergence for the precision we have considered. 
It is also convenient, in order to find the ground-state energy, to start from the hydrogen atom situation
${\bar \gamma}_2={\bar \gamma}_3=0$ where the solution is known, and to crank up 
${\bar \gamma}_2$ and ${\bar \gamma}_3$ progressively. 
In this way, the range where the ground-state energy lies is fairly well known at each stage
of the calculation.

In the practical task of performing the integral on the right-hand side of Eq.(\ref{eqexcSchc}), it is much better
to get rid of the Coulomb interaction term and its singular behaviour. This is done conveniently by performing
the change of variables ${\bf K}'={\bf K}+{\bf Q}$. In this way the integral becomes
\begin{eqnarray}\label{eqchgvar}
\int d{\bf K}' \frac{1}{({\bf K}-{\bf K}')^2}\, a({\bf K}')= \int_{0}^{\infty} dQ  \int_{0}^{\pi} d\theta \sin \theta
 \int_{0}^{2\pi} d\varphi\;a(K_x+Q \sin \theta \cos \varphi,K_y+Q \sin\theta \sin\varphi,K_z+Q \cos \theta)
\end{eqnarray}
where $\theta$ and $\varphi$ are the polar and azimuthal angles for ${\bf Q}$. However, this change of variables
makes it necessary, in order to conveniently perform numerically the integrals, to evaluate
$a(K_x,K_y,K_z)$ for any values of $K_x,K_y$ and $K_z$. This can be done by sampling a number of $K_x, K_y, K_z$
values and infer any $a(K_x,K_y,K_z)$ value by interpolation. In practice, it is rather
$(1+K^2)^2\,a(K_x,K_y,K_z)$ that we have interpolated, since in the case of the hydrogen atom, this function is just a constant
equal to $1$. So, in our exciton case, we do not expect it to have strong variations. This has allowed us to use
a simple three-dimensional linear interpolation. Naturally, we have also restricted the range of variation of our variables by the change of variables $K_{x,y,z}=\tan t_{x,y,z}$, with the regular discretization being on the $t_{x,y,z}$'s.

It is convenient to reduce the range of the ${\bf K}$ variables by making use of symmetries, for example planar
reflexions. If we consider first the change $K_z \to -K_z$, it produces a change of sign for $D$ in Eq.(\ref{eqdefGD}), all the
other terms being unchanged. As a result, one checks easily that, if $a_n(K_x,K_y,K_z)$ is solution of Eq.(\ref{eqexcSchc}),
then $(-1)^{n-1}a_n(K_x,K_y,-K_z)$ is also solution. As above, we can build a solution that is even under this transform
(and is expected to correspond to the ground state). It satisfies $a_n(K_x,K_y,-K_z)=(-1)^{n-1}a_n(K_x,K_y,K_z)$.
Note that the time-reversed solution $a_n^T({\bf K})=(-1)^{n-1}a^*_{5-n}({\bf K})$ transforms as the odd combination in
this transform $a_n^T(K_x,K_y,-K_z)=(-1)^n a_n^T(K_x,K_y,K_z)$.

Similarly, in the change $K_x \to -K_x$, $F$ is changed into $F^*$ and $D$ into $-D^*$. One checks  that, if $a_n(K_x,K_y,K_z)$ 
is solution, then $a_{5-n}(-K_x,K_y,K_z)$ is solution. Hence, we can take a solution with the symmetry $a_n(-K_x,K_y,K_z)=
a_{5-n}(K_x,K_y,K_z)$. Finally $K_y \to -K_y$ changes $F$ into $F^*$ and $D$ into $D^*$. So, if $a_n(K_x,K_y,K_z)$
 is solution, then $(-1)^{n-1} a_{5-n}(K_x,-K_y,K_z)$ is solution, and we can take a solution with the symmetry 
 $a_n(K_x,-K_y,K_z)=(-1)^{n-1} a_{5-n}(K_x,K_y,K_z)$. Note that, with this choice of solution satisfying simultaneously these
 three symmetries, if we perform a symmetry with respect to the origin $K_x \to -K_x , K_y \to -K_y$ and $K_z \to -K_z$, we find
 that our solution is invariant, as we expect from the ground-state wave function. These symmetries allow us to restrict
 the range of our variables to $K_x \ge 0, K_y \ge 0, K_z \ge 0$.
 
 We want to explore the whole range of ${\bar \gamma}_2$ and ${\bar \gamma}_3$ values allowed by the condition Eq.(\ref{eqcondgambar}). 
 However, we note that if we change the sign of both ${\bar \gamma}_2$ and ${\bar \gamma}_3$,
 from Eq.(\ref{eqdefGD}) $\Delta$, $D$ and $F$ change sign, that is, $h_h$ defined by Eq.(\ref{eqHtr}) is changed into $-h_h$.
 But $h_h$ and $-h_h$ must have the same eigenvectors, and moreover have also the same doubly degenerate eigenvalues
 $\pm \lambda$ as seen in Eq.(\ref{eqhh2}) and Eq.(\ref{eqD2F2}). In other words, going from $h_h$ to $-h_h$ 
 amounts to merely relabeling
 the eigenvectors corresponding to the two positive eigenvalues as corresponding to the two negative eigenvalues, and conversely.
 It is clear that such a relabeling, which is merely a basis change, 
 does not change the eigenvalue spectrum of the whole Hamiltonian we are interested in.
 Accordingly, in particular, the exciton ground-state energy is unchanged when changing the sign of both ${\bar \gamma}_2$ and ${\bar \gamma}_3$.
If we have a solution for a given sign of the ${\bar \gamma}$'s, we are able to obtain a solution having the same groud-state energy
for the ${\bar \gamma}$'s with opposite sign.

Let us now consider what happens if we change only the ${\bar \gamma}_2$ sign, ${\bar \gamma}_3$ being unchanged.
From Eq.(\ref{eqdefGD}), one finds that $\Delta$ is changed into $-\Delta$, and $F$ into $-F^*$, while $D$ is unchanged.
Moreover, as we have seen above, in the change of variables $K_x \to K_y$ and $K_y \to -K_x$, 
$D({\bf K})$ becomes $i\,D({\bf K})$, while $F({\bf K})$ becomes $-F({\bf K})$. But we notice that a further simple change
of functions $a_{2} \to -ia_{2}, a_{4} \to -ia_{4}$ is equivalent to changing $D$ into $iD$. Combining these three changes,
we have changed $\Delta$ into $-\Delta $, $F$ into $F^*$ and $D$ into $-D$. But the resulting equations for 
Eq.(\ref{eqexcSchb}), or Eq.(\ref{eqexcSchc}), are identical to the original ones provided we make the additional change
of functions $a_1 \leftrightarrow a_3 , a_2 \leftrightarrow a_4$. Hence, we conclude that ${\bar \gamma}_2 \to
-{\bar \gamma}_2$ leaves the energy spectrum invariant. Combining now the two changes of signs for the $\gamma$'s,
we obtain that the ground-state energy depends only on $|{\bar \gamma}_2|$ and $|{\bar \gamma}_3|$, so we can
restrict our study to $0 \le {\bar \gamma}_2 < 1/2$, $0 \le {\bar \gamma}_3 < 1/2$.

Let us finally examine the limiting cases ${\bar \gamma}_2 \to 1/2, {\bar \gamma}_3 \to 1/2$.
As we have seen below Eq.(\ref{eqmLmH}), when either ${\bar \gamma}_2=1/2$ or ${\bar \gamma}_3=1/2$, 
there are some ${\bf K}$ directions for which $K^2-\lambda^2=0$. As a result, if in Eq.(\ref{eqexcSchc}) ${\bar \epsilon}$
is fixed, for large values of $K$, the denominator in the first factor of the right-hand side will go to zero. This implies that
$a({\bf K})$ will not behave for all ${\bf K}$ directions as $1/K^2$, as we have found in Eq.(\ref{eqexcSchcH}).
This leads to a divergent behaviour in the integral of Eq.(\ref{eqexcSchc}). The natural way to
escape this problem is to let ${\bar \epsilon}$ be very large, so this divergent behaviour of $a({\bf K})$ is pushed to ever higher values
of $K$, and one avoids an actual singularity in the limits ${\bar \gamma}_2 \to 1/2, {\bar \gamma}_3 \to 1/2$.

Unfortunately, an explicit handling of this behaviour is not so easy in general because it arises from the behaviour of $a({\bf K})$ 
around some specific directions of ${\bf K}$. Nevertheless when we let both ${\bar \gamma}_2 \to 1/2, {\bar \gamma}_3 \to 1/2$ at the same time, it is possible to extract the divergent behaviour of ${\bar \epsilon}$ by a simple rescaling. Let us set
$2 \,{\bar \gamma}_2 =1-\eta_2, 2\,{\bar \gamma}_3 =1-\eta_3$, and consider $\eta_2 \to 0, \eta_3 \to 0$ with $r=\eta_3/\eta_2$
fixed to a finite value. We perform the rescaling ${\bf K}={\bar {\bf K}}/\eta_2$, while at the same time setting ${\bar \epsilon}=
\alpha /\eta_2$. This leads, for the denominator on the right-hand side of Eq.(\ref{eqexcSchc}), to 
$(K^2+{\bar \epsilon})^2-\lambda^2=(2/\eta_2^3)[{\bar K}^4+\alpha {\bar K}^2 + 3(r-1)({\bar K}_x^2\,{\bar K}_y^2 +{\bar K}_y^2\,{\bar K}_z^2 +{\bar K}_z^2\,{\bar K}_x^2)]$.
In this rescaling, the numerator provides a factor $1/\eta_2^2$, while another factor $1/\eta_2$ comes from the 
rescaling of the variable ${\bf K}'$ into ${\bar {\bf K}}'$ in the integral. Hence, the scaling factor $\eta_2$ 
disappears completely and we are left with the following equation, free of singularity
\begin{eqnarray}\label{eqexcSchcresc}
b({\bar{\bf K}}) = \frac{{\bar K}^2 \mathbb{1}-{\bar h}_h({\bar {\bf K}})}
{2[\alpha {\bar K}^2+{\bar K}^4+ 3(r-1)({\bar K}_x^2\,{\bar K}_y^2 +{\bar K}_y^2\,{\bar K}_z^2 +{\bar K}_z^2\,{\bar K}_x^2)]}
\frac{1}{\pi^2} \int d{\bar {\bf K}}' \frac{1}{({\bar {\bf K}}-{\bar {\bf K}}')^2}\, b({\bar {\bf K}}')
\end{eqnarray}
where $b({\bar{\bf K}})=a({\bar {\bf K}}/\eta_2)$. We have solved numerically this equation 
for the scaling factor $\alpha$ in the energy,
in the case $r=1$, and we have found $\alpha = 0.43$. This is in fair agreement with the corresponding reduced energy 
${\bar \epsilon}=4.77$ we have found by a direct numerical solution of Eq.(\ref{eqexcSchc}) for $\eta_2=\eta_3=0.1$.

\begin{figure}
\centering
{\includegraphics[width=0.8\linewidth]{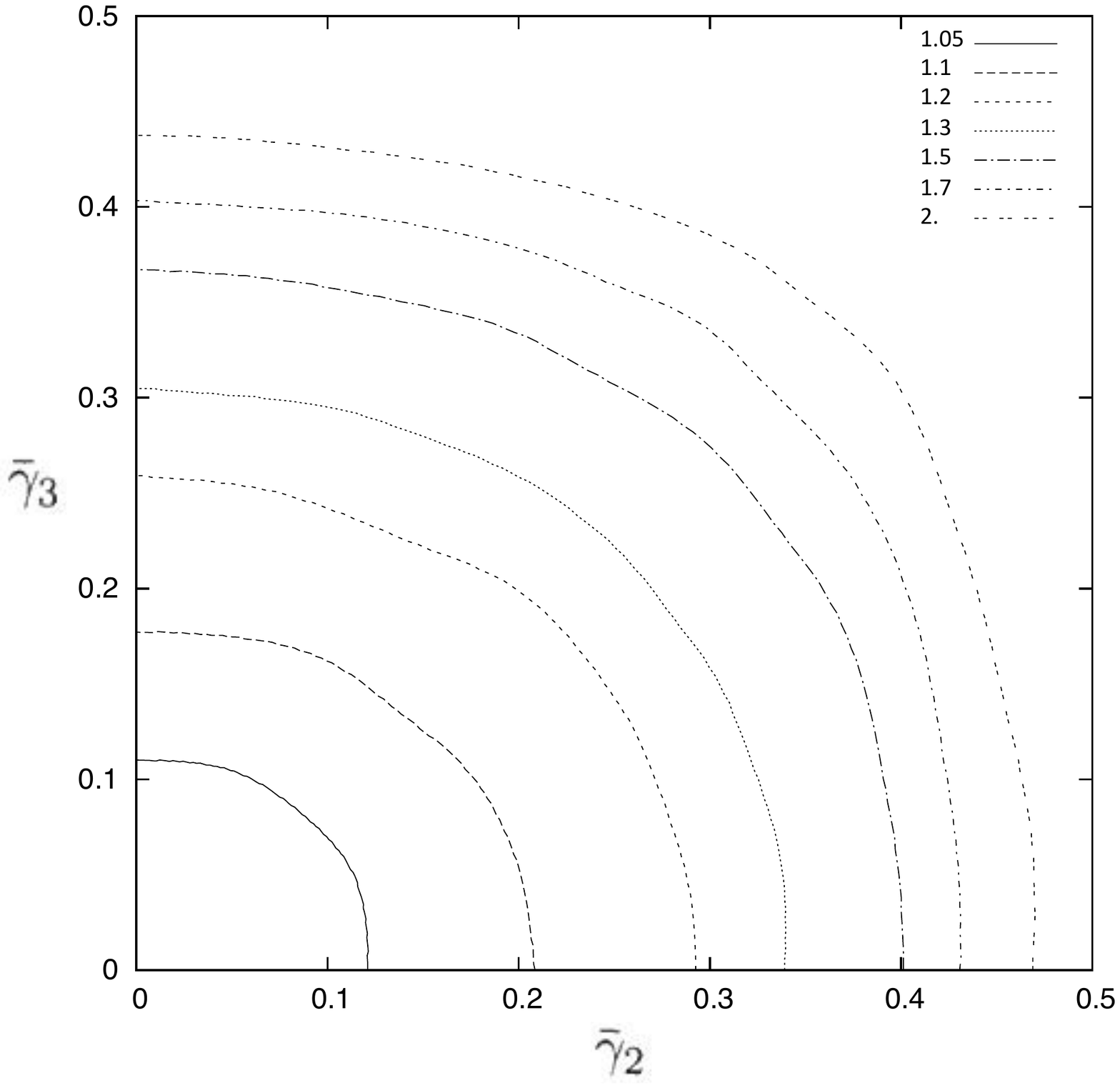}}
\caption{Reduced ground-state exciton binding energy ${\bar \epsilon}$ in the ${\bar \gamma}_2 - {\bar \gamma}_3 $  plane.
The level lines for ${\bar \epsilon}=1.05, 1.1, 1.2, 1.3, 1.5, 1.7$ and $2.$ are represented, as indicated in the figure.}
\label{fig}
\end{figure}

Following the procedure indicated above, we have calculated the reduced exciton ground-state energy ${\bar \epsilon}$
from Eq.(\ref{eqexcSchc}) as a function of ${\bar \gamma}_2 $ and ${\bar \gamma}_3 $, on a grid with spacing $0.05$ for these two
parameters. Rather than presenting these extensive results as a table, we have used them to evaluate ${\bar \epsilon}$ for any value
of ${\bar \gamma}_2 $ and ${\bar \gamma}_3 $ by a two-dimensional spline interpolation, and then obtain the level lines for various
values of ${\bar \epsilon}$ in the ${\bar \gamma}_2 - {\bar \gamma}_3 $ plane. They are displayed in Fig.\ref{fig}. We have chosen not
to smooth out the small irregularities which are apparent along these level lines, since they result from the imprecision
of our calculations and accordingly give a fairly direct information on it. In particular, the results for ${\bar \epsilon}$ 
close to $1$ are very sensitive to this imprecision, and the resulting line for ${\bar \epsilon}=1.05$ would be pretty bad. 
Hence, to draw this line, we have rather relied on a quadratic interpolation from the ${\bar \gamma}_2 =0.1$ and ${\bar \gamma}_3=0.1 $ results, rather than using our ${\bar \gamma}_2 =0.05$ and ${\bar \gamma}_3=0.05 $ results.

The numerical results are fairly regular, with the represented level lines being not so far from quarter circles, provided
we do not go too close to the boundaries ${\bar \gamma}_2 =0.5$ and ${\bar \gamma}_3=0.5 $. This means that,
roughly speaking, ${\bar \epsilon}$ depends only on ${\bar \gamma}_2^2 + {\bar \gamma}_3^2 \equiv 2{\bar \gamma}^2 $.
Correspondingly, we find that ${\bar \epsilon}-1 \simeq 2.5 (1/\sqrt{1-4{\bar \gamma}^2}-1)$ provides an approximate representation
of the numerical results. In particular, it is quadratic for small ${\bar \gamma}$ and it diverges for ${\bar \gamma} \to 1/2$. It gives
${\bar \epsilon} \simeq 4.23 $ for ${\bar \gamma}=0.45$, not so different from the result from our numerical
${\bar \epsilon}=4.77$ mentioned above.

Regarding the size of the results, we note that taking our reduced ground-state exciton binding 
energy as unity ${\bar \epsilon}=1$ corresponds to
making use, in the standard expression $E=- \mu_X (e^2/4\pi \epsilon_{sc})^2/(2 \hbar^2)$, of a reduced mass $\mu_X$
given by $1/\mu_X=1/m_e+1/\mu_h$, where from Eq.(\ref{eqmLmH}) and Eq.(\ref{eqgam1bar}) the reduced hole mass 
$\mu_h$ itself is taken equal to the half-sum of the reduced heavy and light hole masses $1/\mu_h=(1/m_H + 1/m_L)/2$.
This is a frequently done approximation, for lack of better knowledge. Fig.\ref{fig} shows that in principle we may have
a fairly important departure from this simple approximation.

However, we have considered in practice a few semiconductors where reliable data exist for the valence band as well as the
conduction band parameters \cite{cardona,kara,vurga,vurgap}. One can still see sizeable theoretical and experimental uncertainties in the knowledge of these band parameters \cite{kara,vurga}. 
Remarkably, the strong variations among compounds in the basic Luttinger parameters and conduction electron mass, 
provide a much reduced variation in our reduced Luttinger parameters 
${\bar \gamma}_2 $ and ${\bar \gamma}_3 $. We note that a light electronic mass increases ${\bar \gamma}_1 $
from Eq.(\ref{eqgam1bar}), which leads to a reduction of the size of ${\bar \gamma}_2 $ and ${\bar \gamma}_3 $.
We find that ${\bar \gamma}_2 $ is mostly in the range $0.1-0.15$, while ${\bar \gamma}_3 $ takes slightly higher
values in the range $0.1-0.2$. From our results, the warping leads to an increase of the exciton binding energy by about $10 \%$.
This should be taken into account as soon as one looks for some precision in the exciton
binding energy. The largest correction we have found for a compound with specifically known parameters is in the case of BN
\cite{vurga}, where ${\bar \gamma}_2 =0.12$ and ${\bar \gamma}_3 = 0.2 $ leads to a $15 \%$ increase
from the warping in the exciton binding energy. However, it should be kept in mind that this binding energy is quite sensitive to the
parameters values in this range. For example having ${\bar \gamma}_2 ={\bar \gamma}_3 = 0.2 $ would lead to a $20 \%$
increase in the binding energy, and ${\bar \gamma}_2 ={\bar \gamma}_3 = 0.25 $ would give a $35 \%$ increase.
Hence, keeping in mind the uncertainties on the valence band parameters, the effect of warping on the exciton binding energy
may happen to be quite important.

\section{Conclusion}

In this paper, we have addressed the effect on the exciton ground-state energy of the warping of the valence band near its edge. We have specifically considered the case of bulk exciton for semiconductors with zinc-blende crystal structure. 
Assuming as usual that the split-off subband due to spin-orbit coupling is energetically far enough to play a negligible role, 
we have considered the four-dimensional problem for the hole kinetic energy. We have shown that, due to the cubic symmetry,
the exciton ground state has a fourfold degeneracy. 

We have introduced, by a simple scaling, reduced
Luttinger parameters ${\bar \gamma}_2$ and ${\bar \gamma}_3$, whose absolute value are less than $1/2$. We have
studied systematically the exciton ground-state energy as a function of these reduced Luttinger parameters, by solving
numerically the integral equation corresponding to the Schr\"odinger equation in Fourier transform. We find that, compared
to its standard expression in the absence of warping (corresponding to the case ${\bar \gamma}_2={\bar \gamma}_3=0$),
the exciton binding energy can in principle increase without limitations. For moderate increases, we have provided an
approximate analytical expression for the increase in terms of the Luttinger parameters. 
Going through the values of band structure
parameters, and in particular Luttinger parameters, for tabulated semiconductors, we find a typical $10 \% - 15 \%$ increase
in the exciton binding energy. However, this energy is fairly sensitive to the values of the Luttinger parameters, and a slight
increase in their values due to experimental or/and theoretical uncertainties could lead to a markedly larger increase in the
exciton binding energy.

Acknowledgements: The authors are grateful to Monique Combescot for interesting them in
this warping problem and for discussions.


\begin{references}
\bibitem{note1} In terms of those of Ref.\cite{lutkoh}, our coefficients are given by $ \hbar^2 \gamma_1/m_0=2(A+2B)/3$,
$\hbar^2 \gamma_2/m_0=(A-B)/3$ and $\hbar^2 \gamma_3/m_0=C/3$, and we have an additional $\pi /2$ rotation around the $z$ axis.
These $\gamma_i$ are just the coefficients introduced by Luttinger \cite{lutt}.
\bibitem{remdegnum} In the simple case ${\bar \gamma}_3=0$ where this $4 \times 4$ problem reduces to two $2 \times 2$ problems, we have checked numerically this degeneracy (within our numerical precision).
\end{references}
\end{document}